*Article*

# Polarization Calibration of a Microwave Polarimeter with Near-Infrared Up-Conversion for Optical Correlation and Detection

Francisco J. Casas *, Patricio Vielva, R. Belen Barreiro, Enrique Martínez-González and G. Pascual-Cisneros

Instituto de Física de Cantabria (IFCA), Avda. Los Castros s/n, 39005 Santander, Spain
* Correspondence: casas@ifca.unican.es; Tel.: +34-942-200-892; Fax: +34-942-200-935

**Abstract:** This paper presents a polarization calibration method applied to a microwave polarimeter demonstrator based on a near-infrared (NIR) frequency up-conversion stage that allows both optical correlation and signal detection at a wavelength of 1550 nm. The instrument was designed to measure the polarization of cosmic microwave background (CMB) radiation from the sky, obtaining the Stokes parameters of the incoming signal simultaneously, in a frequency range from 10 to 20 GHz. A linearly polarized input signal with a variable polarization angle is used as excitation in the polarimeter calibration setup mounted in the laboratory. The polarimeter systematic errors can be corrected with the proposed calibration procedure, achieving high levels of polarization efficiency (low polarization percentage errors) and low polarization angle errors. The calibration method is based on the fitting of polarization errors by means of sinusoidal functions composed of additive or multiplicative terms. The accuracy of the fitting increases with the number of terms in such a way that the typical error levels required in low-frequency CMB experiments can be achieved with only a few terms in the fitting functions. On the other hand, assuming that the calibration signal is known with the required accuracy, additional terms can be calculated to reach the error levels needed in ultrasensitive B-mode polarization CMB experiments.

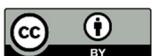

## 1. Introduction

Penzias and Wilson measured in 1964 a noise-like signal [1] that was finally identified as the cosmic microwave background (CMB). This radiation is the remaining footprint of the Big Bang and was postulated by Gamow, Alpher, and Herman in the late 1940s [2]. Many radio astronomy instruments have been used since then to characterize the CMB. Space missions [3–5], as well as balloon-borne (e.g., [6,7]) and ground-based experiments (e.g., [8–13]), have been dedicated to the analysis of temperature and polarization anisotropies of the CMB. These measurements have been an invaluable resource for testing cosmological models and fundamental physics, since the processes that operated in the early universe or acted on the photons during their passage to the Earth have imprinted very weak but distinct features on the otherwise uniform background. The polarization anisotropy patterns are formed by a combination of the electric-like (E) and magnetic-like (B) modes. Recent E-mode polarization measurements confirm the validity of the standard cosmological model [14]. However, primordial B-mode signals have yet to be detected (see the background imaging of cosmic extragalactic polarization (BICEP2) experiment [10], as well as [15–18]). They are fainter and can be easily contaminated, but they may reveal crucial information about the early stages of the universe. Many major questions about inflation, the primordial background of gravitational waves, or the magnetic fields may be resolved by measurements of the B-mode signals. Current and future ground- and space-based [19] CMB polarization experiments are aiming at an unprecedented level of sensitivity. Therefore, systematic effects that were usually considered less important than statistical uncertainties are becoming the most significant limitation at the instrumental level. Such experiments, as many others related to different scientific and technological applications (see, for instance, [20]), need a calibration method providing control over diverse systematic errors, which in the case of CMB polarization experiments are, among the most relevant ones, intensity to polarization leakage, polarization angle, and efficiency errors.

In this work, a polarization calibration [21,22] method is applied to a microwave polarimeter demonstrator based on a near-infrared (NIR) frequency up-conversion. The polarization systematic errors of the demonstrator can be corrected using the proposed calibration technique, providing low polarization percentage and polarization angle errors. Other usual systematic errors of the polarimeter, related to, e.g., the beam, bandwidth, and linearity, should be also calibrated for actual observations, but this work refers only to polarization systematics that can be corrected in the same way in the laboratory and when the instrument is mounted in an observatory (it should be taken into account that, for example, beam calibration depends on the telescope size when the polarimeter is configured to operate in direct imaging).

The proposed methodology assumes the use of a polarized artificial source instead of celestial ones. This has important advantages because the few astronomical candidates suffer from frequency dependence and time variability. Moreover, they are not visible from all observatories and are extended sources. The best option is Tau-A, which allows accuracies for the polarization orientation between 1° and 0.5°, but for ultrasensitive CMB experiments (for instance, Lite satellite for the studies of B-mode polarization and inflation from cosmic background radiation detection (LiteBIRD) [19], CMB stage four (CMB-S4) [23], or probe of inflation and cosmic origins (PICO) [24]) requiring arc-minute-level polarization angle accuracy, the best option is the use of artificial sources that can be extremely well characterized in the laboratory. Another advantage of the signals emitted by artificial sources is that they can be very similar, for instance, in terms of spectral content or shape, to the ones that are observed from the sky when the polarimeters are in their usual operation. On the other hand, although the calibration source used in this work has been applied only to laboratory calibrations, it could also be used (obviously with some modifications) for observatory calibrations, because the signal source can be placed in the near field of the telescope by coupling it at some point of its optical path or even directly in front of the receivers. Due to these reasons, the proposed technique is suitable to be used in many present and future CMB experiments that can be found in the literature (see, for instance, [9,10,20,23–31]). Other methods, such as the one called self-calibration [32], try to overcome the lack of good astronomical calibrators assuming some predictions that prevent the study of some cosmological phenomena, such as cosmic birefringence [33], introducing errors on the cosmological parameters. A good review of previous research about calibration methods for CMB polarization, as well as the advantages and disadvantages of other techniques compared to the based on using artificial sources, can be found in [34–36].

The calibration method proposed here is based on the fitting of polarization percentage and polarization angle errors using sinusoidal functions composed of either multiplicative or additive terms, respectively. The error fitting accuracy increases with the number of terms. As a consequence, assuming that the calibration signal can be characterized and known with a low enough uncertainty, and that the instruments operate in stable conditions to avoid the need of further calibration measurements, it is shown that the typical systematic error requirements of low-frequency CMB experiments such as QUIJOTE (Q–U–I joint Tenerife experiment) [25–29] or LSPE-Strip (strip instrument of the large scale polarization explorer) [30] (around 0.5° in polarization angle error and 1% in polarization percentage error), can be reached generally using error functions with only one or two terms. Additionally, the systematic error-level requirements of highly sensitive future experiments [19,23,24] can be reached by adding more terms to the fitting functions.

This work is organized as follows: the experimental setup is described in Section 2; a description of the proposed calibration technique is presented in Section 3; in Section 4 the technique is applied to laboratory measurements of the polarimeter demonstrator providing some representative examples; Section 5 is a discussion about the required number of terms in the fitting functions; lastly, Section 6 draws general conclusions.

## 2. Experimental Setup

In this work, the proposed polarization calibration method is applied to a polarimeter demonstrator that was already presented in a previous study [37] and allows optical correlation and signal detection at a wavelength of 1550 nm. Figure 1 shows a simplified block diagram of the polarimeter.

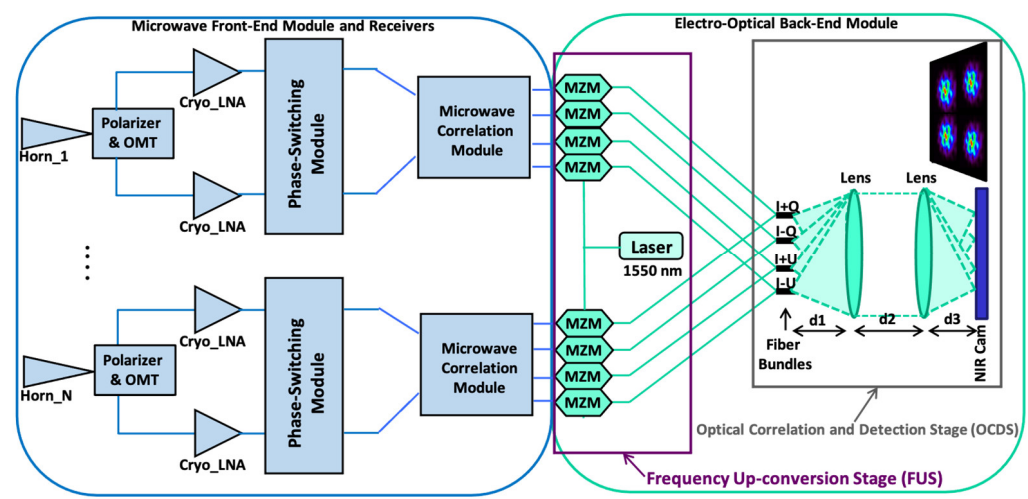

**Figure 1.** Simplified block diagram of the polarimeter demonstrator with correlation/detection in the near-infrared (1550 nm). Reproduced with permission from F. J. Casas, Sensors; published by MDPI, 2019 [37].

Each receiver of the polarimeter has four NIR output signals that are modulated by means of a microwave phase-switching module. That modulation allows the measurement of the polarization degree and the polarization angle [27,37] from each one of the output signals independently.

The prototype mounted in the laboratory is composed of a front-end module (FEM) connected to two microwave receivers, operating from 10 to 20 GHz, and an electro-optical back-end module (EOBEM) with a frequency up-conversion stage (FUS) at the input, connected to an optical correlation and detection stage (OCDS). It represents a solution for the implementation of ultrasensitive large-format interferometers to measure the polarization B-modes at the lowest frequencies of the CMB spectrum. The polarimeter was designed to measure the polarization of the microwave radiation from the sky, obtaining the I, Q, and U Stokes polarization parameters [38] of the incoming signal simultaneously, in a frequency range from 10 to 20 GHz. The microwave receivers of the polarimeter share the conceptual design of those of QUIJOTE experiment's 30 and 40 GHz instruments (TGI and FGI); thus, the proposed methodology can also be applied to those cases. The detection stage is composed by the EOBEM with input microwave signals entering the NIR FUS, composed of a laser and a set of commercial Mach–Zehnder modulators (MZM), as well as an OCDS implemented basically with a fiber array, a pair of lenses, and a camera.

An advantage of this concept is that the same OCDS can be used to operate both as a synthesized-image interferometer, such as the Q–U bolometric interferometer for cosmology (QUBIC) [31], and as a traditional imager, only by changing the distances among the fiber array, the lenses, and the camera ($d_1$, $d_2$, and $d_3$ in Figure 1). To operate as a synthesized image interferometer, a $6f$ optical configuration is used, where $f$ is the focal length of the lenses. In this case, $d_1 = 2f$, $d_2 = 3f$, and $d_3 = f$. On the other hand, to operate as an imaging instrument, a $4f$ optical configuration is used, with $d_1 = d_3 = f$ and $d_2 = 2f$. In the first case, the instrument provides a synthesized image of the polarization parameters of the microwave radiation coming from the sky. In the second case, the OCDS is basically a NIR detection stage of the up-converted signals that have the polarization information of the microwave radiation. In the latter case, a telescope is required to focus the signal from the sky to the instrument.

*Polarization Calibration Laboratory Test-Bench*

A sketch of the calibration test-bench mounted in the laboratory is shown in Figure 2. The kind of measurements performed, and the calibration setup are also described in [37].

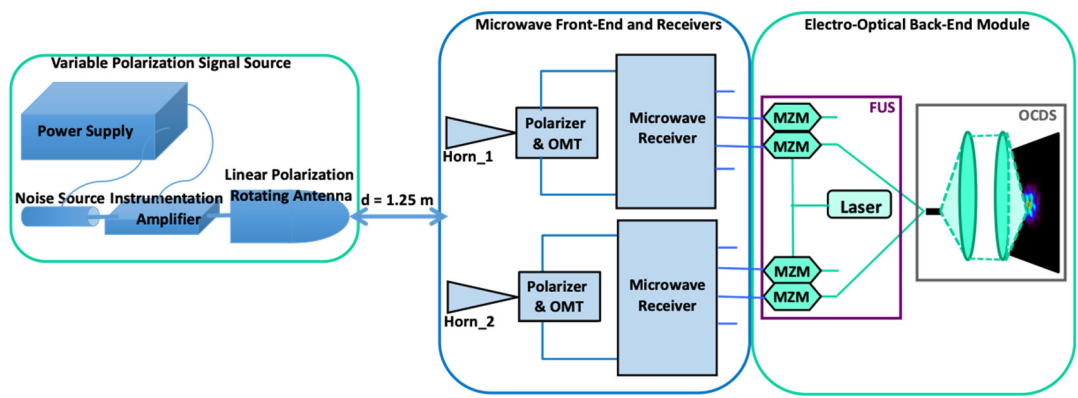

**Figure 2.** Sketch of the polarimeter demonstrator's calibration test-bench. Reproduced with permission from F. J. Casas, Sensors; published by MDPI, 2019 [37].

The calibration source implemented in the laboratory is an external microwave signal source with variable polarization angle, implemented by means of a rotating antenna connected to a wideband amplifier and a noise source (for more details, see [37]). It provides a wideband (10–20 GHz) linearly polarized input signal with a variable hand-controlled polarization angle. The calibration signal presents a polarization angle uncertainty of about 1° due to the source's simple polarization hand-controlled mechanical system. On the other hand, in terms of polarization purity, the cross-polarization provided by the source is approximately −30 dB. In order to achieve the arc-minute-level polarization errors required in actual CMB polarization experiments, more sophisticated sources, presenting electronically-controlled polarization angles, lower cross-polarization components, and carefully aligned calibration setups [34–36], should be used; however, for the calibration method demonstration, it is possible to use the reported source considering that for an actual calibration the emitted signal must be known with the required accuracy and error levels of that particular experiment.

## 3. Calibration Technique

In this section, a polarization calibration methodology for the microwave polarimeter demonstrator of Figure 1 is described. This work focuses on the characterization of the polarization percentage (or efficiency) and polarization angle errors from the instrument's modulated output signals, when a known microwave-polarized signal is measured. Analytical expressions of these errors are obtained to remove them directly from the measured values. As the correct polarization angle is unknown when sky observations are taken, the error expressions are given as a function of the measured polarization angle and not as a function of the calibration signal polarization angle as would, in principle, be the natural way to determine them.

The previously mentioned external broadband (10–20 GHz) noise source provides a totally linearly polarized (100%) signal with a variable polarization angle from 0° to 157.5°, using incremental steps of 22.5° to cover the complete polarization cycle (polarization angles of 180° and 0° are equivalent). Smaller incremental steps could be also used (for instance, 10° or a lower quantity) resulting in a higher number of measurements required to cover the complete polarization cycle and, consequently, a more accurate characterization of the systematic errors. However, 22.5° is chosen here for simplicity. On the other hand, the power of the source was previously calibrated to assure a linear operation of the receivers.

As said above, the polarimeter demonstrator has four NIR output signals per receiver, when operating in direct imaging mode, or in total, when operating as a synthetized imaging interferometer (Figure 1). In the first case, each receiver must be calibrated separately, while, in the second, the overall instrument can be calibrated using the same methodology. In any case, the output signals are modulated by means of electrical phase-switching modules that introduce a phase shift sequence of 0°, 90°, 180°, and 270° between the two branches of the receivers. That modulation affects the polarization of the incoming signals, allowing the measurement of the polarization efficiency (or percentage) and the polarization angle. From those output signals, it is also possible to characterize the instrumentation systematic errors that are also modulated, so that they can be fitted by means of sinusoidal functions.

*3.1. Polarization Angle Calibration*

The calibration source emits broadband polarized signals during a number of measurements ($N_m$) equal to eight ([0°, 22.5°, 45°, …157.5°] of input polarization angles ($\alpha_{in}$)). Then, from each one of the four polarimeter outputs, the corresponding measured polarization angles ($\alpha_m$) are obtained, and it is possible to define measured polarization angle errors ($\varepsilon_{\alpha m}$) arrays as

$$\varepsilon_{\alpha m} = \alpha_{in} - \alpha_m. \quad (1)$$

To correct the measured polarization angle, an error function is defined in such a way that it fits the $\varepsilon_{\alpha m}$ arrays. This error function is called fitted polarization angle error ($\varepsilon_{\alpha f}$). Considering the $\varepsilon_{\alpha f}$ error function, the corrected polarization angle ($\alpha_c$) can be defined as

$$\alpha_c = \alpha_m + \varepsilon_{\alpha f}, \quad (2)$$

which also applies to each of the polarimeter outputs. The following step is then to define the $\varepsilon_{\alpha f}$ functions that are implemented here as the sum of $n$ terms with $n$ high enough to fit the $\varepsilon_{\alpha m}$ with the required accuracy,

$$\varepsilon_{\alpha f} = \sum_{i=1}^{n} \varepsilon_{\alpha f\_i}, \quad (3)$$

with index $i$ going from 1 to $n$. One advantage of using the reported method is that it is not limited to a predefined number of terms, in such a way that it is possible to get as little error as desired in the fitting process. It is important to note that $\varepsilon_{\alpha f}$ is defined as a function of $\alpha_m$ instead of $\alpha_{in}$ because it is unknown when the instrument is in regular operation (in fact, it is one of the main observables to be obtained). As shown in the next section, both polarization angle and percentage errors have sinusoidal shapes; thus, we can fit them using the following expressions for the $\varepsilon_{\alpha f\_i}$ terms:

$$\varepsilon_{\alpha f\_i} = \varepsilon_{m\_i} + 2[A_i \cos(2\varphi_i - \gamma_i)]. \quad (4)$$

All the parameters of Equation (4) are functions of the measured polarization angle array $\alpha_m$ and, taking into account that, in the calibration measurements, $N_m$ input polarization angles are used, they can be defined as shown below.

$$\varepsilon_{m\_i} = \frac{1}{N_m}\sum_{j=1}^{N_m} \varepsilon_{\alpha r\_ij}. \tag{5}$$

$$\varepsilon_{\alpha r\_i+1} = \varepsilon_{\alpha m} - \sum \varepsilon_{\alpha f\_i} = \varepsilon_{\alpha r\_i} - \varepsilon_{\alpha f\_i}.$$

$$\varepsilon_{\alpha r\_1} = \varepsilon_{\alpha m} \text{ (first term calculation)}. \tag{6}$$

The $\varepsilon_{m\_i}$ parameter is the mean value of the remaining polarization angle error ($\varepsilon_{\alpha r\_i}$, Equation (5)) that results when removing the sum of the $\varepsilon_{\alpha f\_i}$ error terms to $\varepsilon_{\alpha m}$ (Equation (6)). For the first term calculation, the remaining error is the measured one; for the second term, the remaining error can also be calculated as the difference between the remaining error from the previous term and the fitted one.

$$\varphi_i = K_i\, \alpha_m. \tag{7}$$

The $\varphi_i$ parameter is a representation of a "harmonic" frequency of the sinusoidal function that can be obtained by the multiplication of the measured polarization angle array and $K_i$ (Equation (7)), which is a real number representing that "harmonic" of the fitting function. For the particular calibration measurements presented in this work, $K_i$ is optimized to minimize $\varepsilon_{\alpha r}$ taking values between 0 and $N_m/2$ because it is supposed a slow (low frequency) evolution of the polarization angle error. In case of presenting a rapid (high frequency) evolution, $K_i$ can take higher maximum values ($N_m$, $2N_m$, …) to accurately fit these fast variations of the polarization angle error.

$$A_i = \sqrt{\text{Re}_i^2 + \text{Im}_i^2}. \tag{8}$$

$$\gamma_i = \tan^{-1}\left(\frac{\text{Im}_i}{\text{Re}_i}\right). \tag{9}$$

$$\text{Re}_i = \frac{1}{N_m}\sum_{j=1}^{N_m}\left(\varepsilon_{\alpha r_{ij}} - \varepsilon_{m_i}\right)\cos(K_i\cdot\alpha_{m\_j}). \tag{10}$$

$$\text{Im}_i = \frac{1}{N_m}\sum_{j=1}^{N_m}\left(\varepsilon_{\alpha r\_ij} - \varepsilon_{m\_i}\right)\sin(K_i\cdot\alpha_{m\_j}). \tag{11}$$

The $A_i$ and $\gamma_i$ parameters are the amplitude and phase of the sinusoidal function that is used to fit the error. They are achieved from the real and imaginary components (see Equations (10) and (11)) that are calculated from $K_i$, $\varepsilon_{m\_i}$, and the elements of the $\varepsilon_{\alpha r}$ and $\alpha_m$ arrays.

The reported fitting method can provide a better matching of the error by employing a higher number (*n*) of $\varepsilon_{\alpha f\_i}$ terms; hence, the logical way to apply it would be to calculate terms of the error function until reaching the required $\varepsilon_{\alpha r}$ level for that particular experiment, or until verifying that the addition of more terms does not appreciably reduce the $\varepsilon_{\alpha r}$. The number of terms required is different depending on the shape of the errors. In the case shown in this work, it was verified empirically that, using one or two terms, the maximum values of $\varepsilon_{\alpha r}$ are around 0.5°, and, using four or five terms, the maximum values of $\varepsilon_{\alpha r}$ are around 0.1°, with respect to the $\alpha_{in}$ actual values. Initially, it was also considered that the calibration signal is known with an uncertainty significantly lower than these values (negligible in an ideal case), in such a way that the fitting process determines the final error. An advantage of this method is the simplicity of the overall analytical expression for the fitted error, which is a simple sum of error terms. This fact significantly reduces the computational requirements when applying it to long-term CMB observations.

## 3.2. Polarization Eficiency Calibration

On the other hand, to calibrate the polarization percentage or efficiency, a similar process is also proposed but presenting some modifications. In this case, from each one of the four polarimeter outputs, the corresponding measured polarization percentages ($p^P_m$) are obtained, and it is possible to define measured polarization percentage error ($\varepsilon p^P_m$) arrays as

$$\varepsilon p^P_m = p^P_S - p^P_m, \tag{12}$$

where $p^P_S$ is the polarization degree of the calibration signal, which, generally, is considered equal to 1 (100% polarized). This can be assumed in a practical case with very low errors (about 0.1% in the case of the calibration source referred in this work), due to the characteristics of the waveguide circuits and antennas generally used to generate the calibration signals.

To correct the polarization efficiency, a multiplicative error function (fitted polarization percentage correction factor or $fp^P_f$) is introduced in such a way that the corrected polarization percentage ($p^P_c$) can be expressed as

$$p^P_c = p^P_m \, fp^P_f. \tag{13}$$

The $fp^P_f$ parameter must fit the measured polarization percentage correction factor ($fp^P_m$) that is defined here as

$$fp^P_m = \left. p^P_S \middle/ p^P_m \right.. \tag{14}$$

Equations (12) and (13) represent the main difference with the polarization angle correction method because for the polarization efficiency the natural form of the fitting function is a product of $n$ terms with $n$ high enough to fit the $fp^P_m$ parameter with the required accuracy:

$$fp^P_f = \prod_{i=1}^{n} fp^P_{f\_i}, \tag{15}$$

where index $i$ goes from 1 to $n$. The advantage of not being limited to a predefined number of terms, in such a way that it is possible to get as little error as desired in the fitting process, is also given in this case. Again, the $fp^P_f$ parameters are defined as functions of the $\alpha_m$ array, and it is possible to define them by using the following expression for the $fp^P_{f\_i}$ terms:

$$fp^P_{f\_i} = f_{m\_i} + 2[A_i \cos(2\varphi_i - \gamma_i)]. \tag{16}$$

All parameters of Equation (16) are functions of $\alpha_m$, and, considering that in the calibration measurements $N_m$ input polarization angles are used, analogously to the polarization angle error case, they can be defined as shown below.

$$f_{m\_i} = \frac{1}{N_m} \sum_{j=1}^{N_m} fp^P_{r\_ij}. \tag{17}$$

$$fp^P_{r\_i+1} = \left. p^P_S \middle/ p^P_{c\_i} \right. = \left. p^P_S \middle/ p^P_m \prod fp^P_{f\_i} \right.. \tag{18}$$

$$fp^P_{r\_1} = fp^P_m \text{ (first term calculation)}.$$

The $f_{m\_i}$ parameter is the mean value of the remaining polarization percentage correction factor ($fp^P_{r\_i}$, Equation (17)) that results when applying the product of the previous $fp^P_{f\_i}$ error terms to $p^P_c$ (Equation (18)). For the first term calculation, the remaining correction factor is the measured one; for the second term, the remaining correction factor can be calculated as the ratio between the polarization degree of the source and the corrected polarization percentage achieved from the previous terms.

$$\varphi_i = K_i\, \alpha_m. \tag{19}$$

The $\varphi_i$ parameter is again a representation of a "harmonic" frequency of the sinusoidal function that can be obtained by the multiplication of the measured polarization angle array and $K_i$, which is a real number representing that "harmonic" of the fitting function. Here also, $K_i$ is optimized to minimize ($fp^{P_r} - 1$) taking values between 0 and $N_m/2$ because it is supposed a low-frequency evolution of the polarization percentage error. However, in cases where a higher-frequency evolution of that error is observed, $K_i$ can take higher maximum values to fit accurately the variations of the $\varepsilon p^P{}_m$.

$$A_i = \sqrt{\mathrm{Re}_i^2 + \mathrm{Im}_i^2}. \tag{20}$$

$$\gamma_i = \tan^{-1}\left(\frac{\mathrm{Im}_i}{\mathrm{Re}_i}\right). \tag{21}$$

$$\mathrm{Re}_i = \frac{1}{N_m} \sum_{j=1}^{N_m} \left(fp^P_{r\_ij} - f_{m\_i}\right) \cos(K_i\, \alpha_{m\_j}). \tag{22}$$

$$\mathrm{Im}_i = \frac{1}{N_m} \sum_{j=1}^{N_m} \left(fp^P_{r\_ij} - f_{m\_i}\right) \sin(K_i\, \alpha_{m\_j}). \tag{23}$$

The $A_i$ and $\gamma_i$ parameters are also the amplitude and phase of the sinusoidal function that is used to fit the correction factor. They are achieved from the real and imaginary components, in Equations (22) and (23), which are calculated from $K_i$, $f_{m\_i}$ and from the elements of the $fp^{P_r}$ and $\alpha_m$ arrays.

As for the polarization angle case, the reported fitting method can provide a better matching of the correction factors by employing a higher number (*n*) of terms in the $fp^{P_f}$ functions; thus, the logical way to apply it would be to calculate terms of the $fp^{P_f}$ until reaching the required polarization percentage error level for each particular experiment, or until verifying that the multiplication of more terms does not reduce appreciably that error. It is important to note that, although the multiplying error functions will increase the noise of the measurement, the signal-to-noise (S/N) ratio will not change as both are multiplied by the correction factors. On the other hand, it is expected that, for actual CMB experiments, the instrumentation will provide $p^P{}_m$ levels of about 0.9 in normal conditions; hence, the correction functions should not increase the noise appreciably.

## 4. Polarimeter Demonstrator Calibration

In this section, the proposed methodology is applied to the polarization angle and polarization percentage errors measured in the direct imaging configuration of the polarimeter demonstrator (4*f* optical configuration in Figure 1). Although the results are not included in this work, the method is equally applicable to the synthetized image configuration measurements (6*f* optical configuration in Figure 1), showing similar performance. As previously commented, the calibration test-bench (see Figure 2) presents an external calibration source composed of a rotating antenna connected to a wideband amplifier and a noise source, the polarization angle of the calibration signal varies from 0° to 157.5° using incremental steps of 22.5° ($N_m$ = 8), and the output signals are electrically modulated in order to measure the polarization percentage (or efficiency) and the polarization angle. Despite the measurement uncertainty limitation provided by the calibration source, the reported method is initially applied considering a calibration signal with a negligible uncertainty, in such a way that the fitting process determines the final error. However, in the next section, a more realistic situation is considered, in which the fitting is performed over the mean values of successive measurement results, until the required uncertainty level is obtained over the measured data to be fitted.

*4.1. Polarization Angle Calibration*

The errors of the four NIR output signals of the polarimeter were fitted following Equations (3)–(11). Five additive error terms $\varepsilon_{af\_i}$ were used for each one of the polarimeter outputs. Figure 3 shows the superposition of the measured errors and those corresponding to the fitted error functions. As can be observed, the errors are given as a function of the measured polarization angles from each output signal.

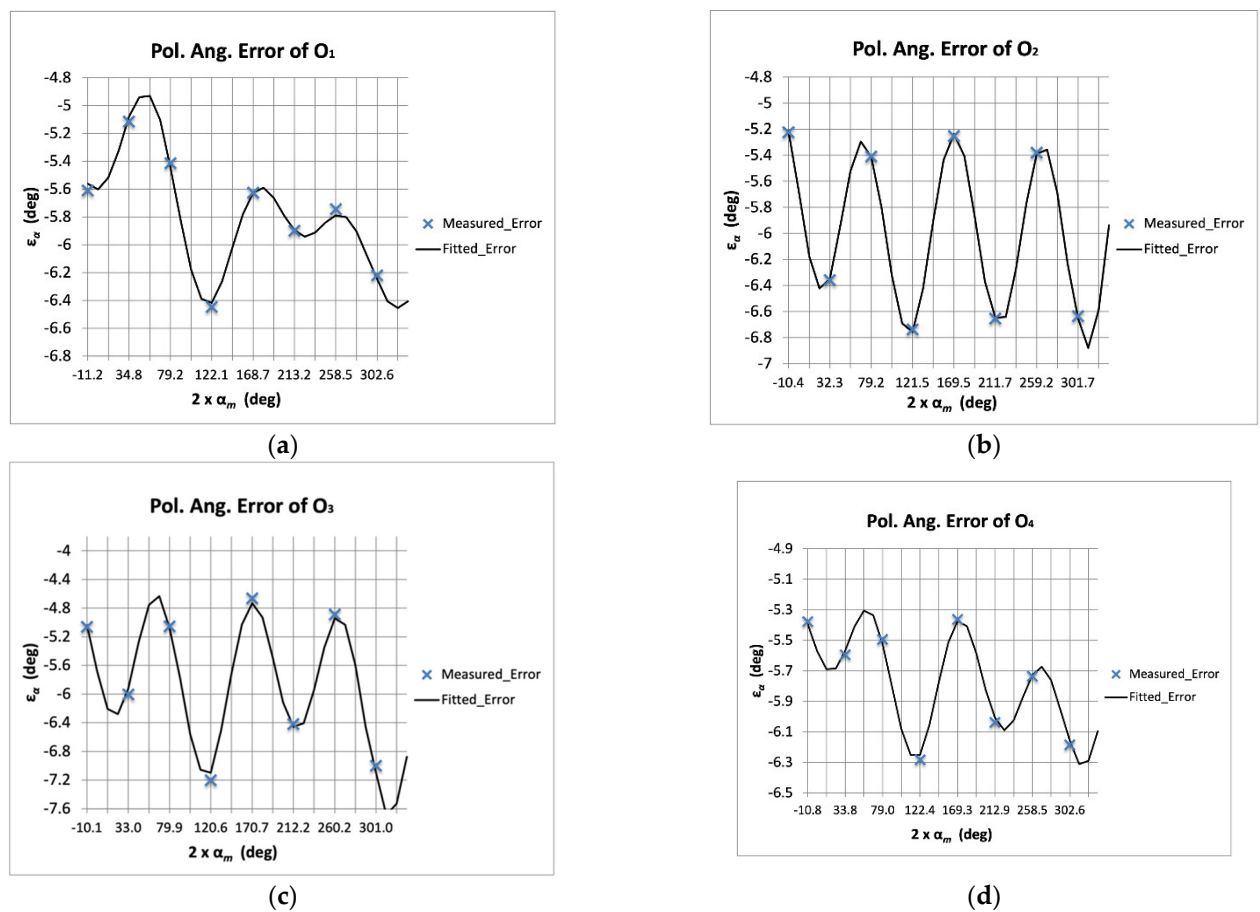

**Figure 3.** Measured polarization angle errors ($\varepsilon_{\alpha m}$, blue crosses) superposed to the fitted ones ($\varepsilon_{\alpha f}$, black line) with five additive terms, given as a function of the measured polarization angles ($\alpha_m$) and for each one of the polarimeter outputs: $O_1$ (**a**), $O_2$ (**b**), $O_3$ (**c**), and $O_4$ (**d**).

Figure 4 shows the comparison between the polarization angle measured errors and those achieved after the application of the error functions (corrected errors). For this particular case, it can be seen that all the outputs present similar remaining error after the application of five additive terms to the error functions.

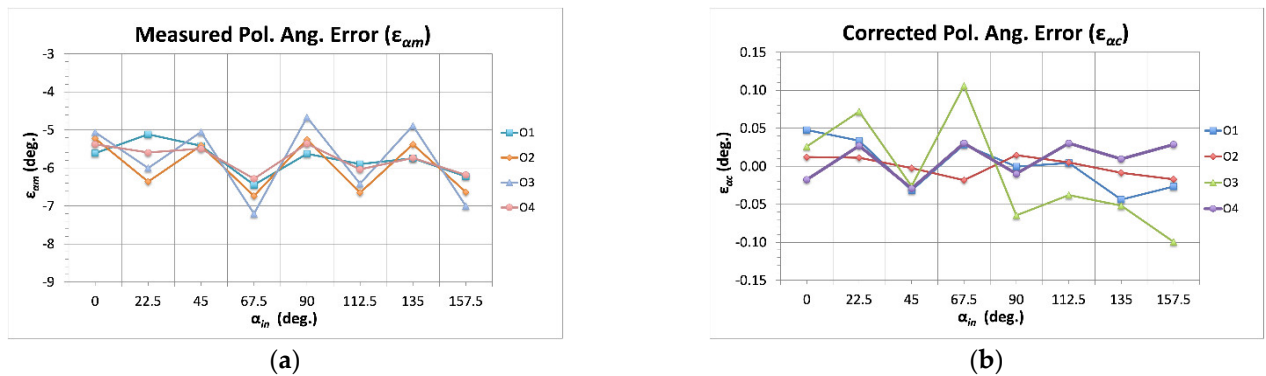

**Figure 4.** Measured polarization angle errors ($\varepsilon_{\alpha m}$, (**a**)) compared to the corrected ones ($\varepsilon_{\alpha c}$, (**b**)). The calibrated errors are represented here as a function of the input polarization angles ($\alpha_{in}$).

Tables 1 and 2 show, respectively, the values of fitting constants $K_i$ (see Equations (7), (10) and (11)) achieved for each one of the five error terms and the maximum remaining polarization angle errors after the application of the fifth error term, for the polarimeter output signals. It can be observed that the polarization angle error was reduced from more than 7° to 0.1° in the worst case ($O_3$). Considering the corrected error values, it is possible to extract the corresponding Q/U isolation worst-case value [27], which is −27 dB (0.1°). This value is five times better than the required isolation (−20 dB) that is usually considered in experiments such as QUIJOTE (maximum polarization angle error of about 0.5°).

**Table 1.** Fitting parameters $K_i$ for each error term and polarimeter output.

| Index $i$ | $K_i$ ($O_1$) | $K_i$ ($O_2$) | $K_i$ ($O_3$) | $K_i$ ($O_4$) |
|---|---|---|---|---|
| 1 | 1.6 | 3.6 | 3.6 | 3.5 |
| 2 | 0.3 | 3.2 | 1.6 | 2.0 |
| 3 | 3.4 | 2.0 | 0.7 | 0.5 |
| 4 | 2.6 | 1.0 | 3.6 | 2.2 |
| 5 | 0.6 | 3.1 | 2.2 | 1.3 |

**Table 2.** Maximum remaining errors after calibration using five error terms.

| Maximum $\varepsilon_{\alpha r}$ | $O_1$ | $O_2$ | $O_3$ | $O_4$ |
|---|---|---|---|---|
| $\varepsilon_{\alpha r\_max}$ (°) | 0.05 | 0.02 | 0.1 | 0.03 |

### 4.2. Polarization Percentage Calibration

For the polarization percentage, it has been considered that the calibration signal is completely polarized ($p^p{}_S = 1$). The errors of the four NIR output signals of the polarimeter were fitted following Equations (16)–(23). Five multiplicative error terms $fp^p{}_{f\_i}$ were used for each one of the polarimeter outputs. Figure 5 shows the superposition of the measured correction factors (Equation (14)) and the fitted ones (Equation (15)). As can be observed, the correction factors are given again as a function of the measured polarization angles from each output signal.

Figure 6 shows the comparison between the polarization percentage measured errors and the achieved after the application of the multiplicative correction factors (corrected errors). For this case, all the outputs present similar remaining error after the calibration process.

Tables 3 and 4 show the values of fitting parameters $K_i$ (see Equations (19), (22) and (23)) achieved for each one of the five error terms and the maximum remaining polarization percentage errors after the application of the fifth term, respectively.

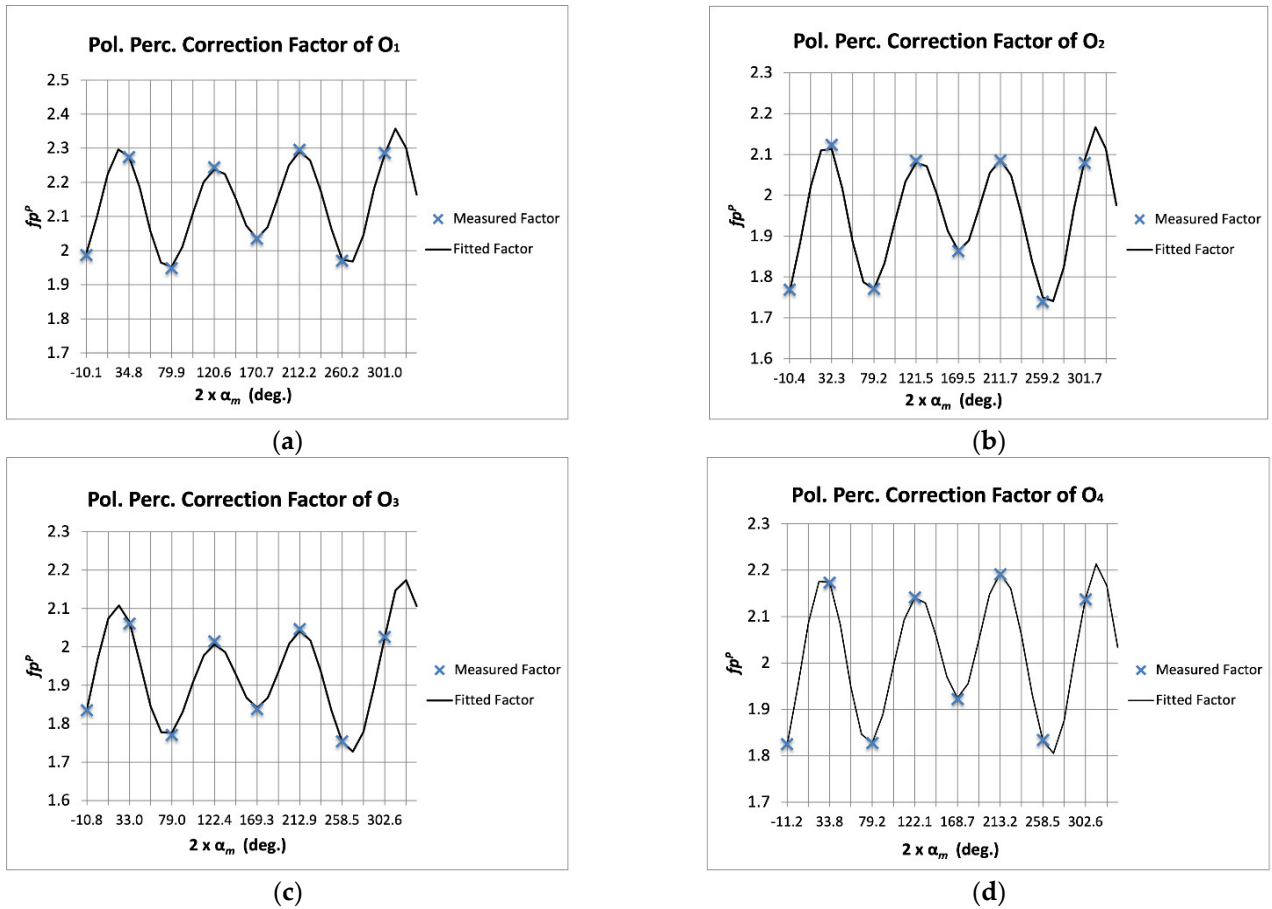

**Figure 5.** Measured polarization percentage correction factors ($fp^P{}_m$, blue crosses) superposed to the fitted ones ($fp^P{}_f$, black traces) with five multiplicative terms, given as a function of the measured polarization angles ($\alpha_m$) and for each one of the polarimeter outputs: $O_1$ (**a**), $O_2$ (**b**), $O_3$ (**c**), and $O_4$ (**d**).

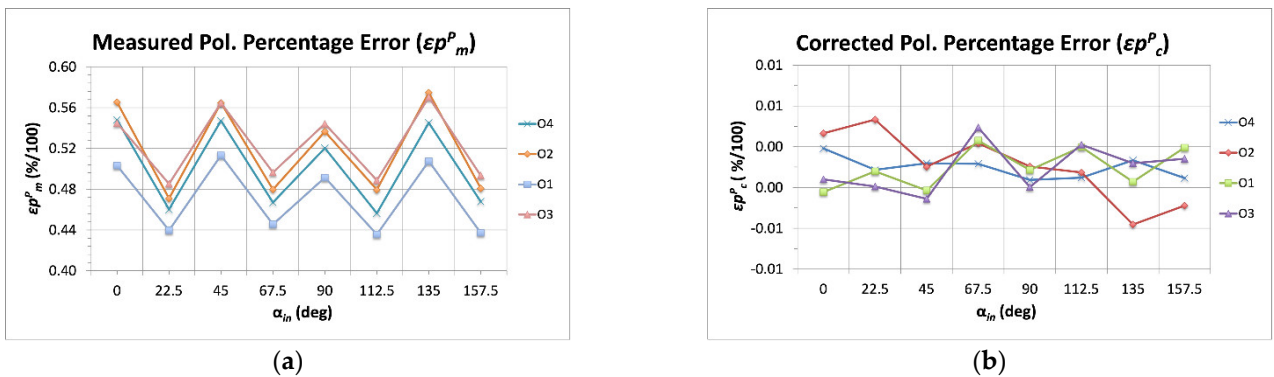

**Figure 6.** Measured polarization percentage error ($\varepsilon p^P{}_m$, (**a**)) compared to the corrected ones ($\varepsilon p^P{}_c$, (**b**)). The calibrated errors are represented here as a function of the input polarization angles ($\alpha_{in}$).

**Table 3.** Fitting parameters $K_i$ for each error term and polarimeter output.

| Index *i* | $K_i$ ($O_1$) | $K_i$ ($O_2$) | $K_i$ ($O_3$) | $K_i$ ($O_4$) |
| --- | --- | --- | --- | --- |
| 1 | 3.6 | 3.6 | 3.6 | 3.5 |
| 2 | 3.0 | 3.0 | 3.1 | 2.6 |
| 3 | 1.3 | 1.4 | 1.1 | 1.6 |
| 4 | 1.8 | 2.5 | 2.2 | 3.1 |
| 5 | 3.5 | 3.6 | 3.0 | 0.4 |

**Table 4.** Maximum remaining error after calibration using five error terms.

| Maximum $\varepsilon p^P$ | $O_1$ | $O_2$ | $O_3$ | $O_4$ |
|---|---|---|---|---|
| $\varepsilon p^P_{max}$ (%) | 0.18 | 0.56 | 0.26 | 0.39 |

It can be observed that the polarization percentage error was reduced from more than 50% to 0.56% in the worst-case output, which can be considered adequate again for experiments such as QUIJOTE, while, for others presenting much more sensitivity [19,23,24], one or two additional error terms would be required.

## 5. Discussion

In this section, the fitting functions number of terms that can be required in an actual case are discussed. Table 5 shows the polarization angle maximum remaining fitting errors after applying each one of the additive terms to the error functions. It can be observed that the first term reduced the error by a factor of ten, while the error was reduced by around 50% (factor of two) with the application of each one of the additional terms. However, looking in detail, there are cases in which the reduction was either lower or higher. For instance, $O_1$ presented an error reduction of only a 7% (from 0.41° to 0.38°) when comparing the use of one and two terms in the fitted error function. Furthermore, $O_3$ presented an error reduction of 22% (from 0.41° to 0.32°) when comparing the use of two and three terms in the fitted error function. On the other hand, $O_2$ and $O_4$ showed a ~50% reduction with the addition of some term (and even higher in some cases) in such a way that the final error was slightly lower. This behavior depends on how well the remaining errors follow a sinusoidal shape. As the $\varepsilon_{af}$ functions are sinusoidal, they obviously optimally fit sinusoidal shapes. In principle, the measured error should present this kind of profile; however, some nonidealities in the measurement test-bench or in the calibration source, or even some non-considered measurement issue could provide this type of result. One example of this can be observed in the second point of the measured error of $O_1$ (see Figure 3) that presented a different trend from the rest of outputs.

**Table 5.** Maximum polarization angle remaining errors $\varepsilon_{ar}$ after the application of each additive term to the $\varepsilon_{af}$ functions.

| N. of Terms | $\varepsilon_{ar\_max}$ (°, $O_1$) | $\varepsilon_{ar\_max}$ (°, $O_2$) | $\varepsilon_{ar\_max}$ (°, $O_3$) | $\varepsilon_{ar\_max}$ (°, $O_4$) |
|---|---|---|---|---|
| 1 | 0.41 | 0.41 | 0.60 | 0.38 |
| 2 | 0.38 | 0.25 | 0.41 | 0.21 |
| 3 | 0.30 | 0.14 | 0.32 | 0.09 |
| 4 | 0.18 | 0.04 | 0.15 | 0.06 |
| 5 | 0.05 | 0.02 | 0.10 | 0.03 |

On the other hand, the number of terms to be applied depends on the requirements of each experiment and the corresponding accuracy of the calibration measurements. As explained in the previous section, there are experiments, with polarization angle error requirements of about 0.5°, which need only one or two terms. However, other experiments presenting higher sensitivity and, consequently, error requirements (for instance, on the order of 1 arc-minute [19]) would require five or more terms. Taking the maximum polarization angle error of the four outputs ($\varepsilon_{ar\_max}$) as an overall reference for the polarimeter, Figure 7 shows its evolution while adding terms to the $\varepsilon_{af}$ function. It can be observed that, from one to four terms, the reduction in the maximum remaining error is quite slow, not reaching the previously quoted 50% reduction factor. However, from four to 10 terms, the reduction factor reaches almost perfectly 50% (higher in some cases), allowing to reach the fitting error level required by any experiment.

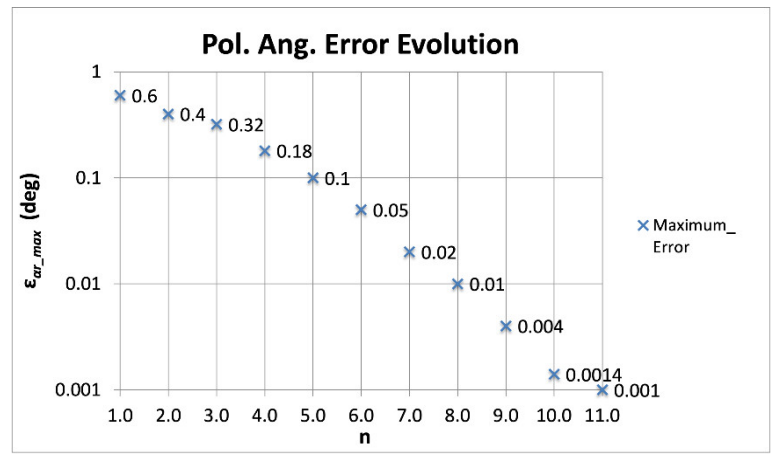

**Figure 7.** Evolution of the maximum polarization angle error ($\varepsilon_{ar\_max}$) as a function of the number of terms (*n*) of $\varepsilon_{af}$, considering the four outputs of the polarimeter.

The previous argumentation considers calibration measurement results with uncertainty levels much lower (ideally negligible) than the remaining fitting error values, in such a way that the fitting process determines the final calibration error. For an actual case in which the calibration setup provides a given uncertainty (for instance, the calibration source described in this work provides an accuracy of 1°), the practical procedure consists of fitting the errors with the required number of terms to have a remaining fitting error significantly lower (for instance, 10 times lower) than the uncertainty provided by the calibration setup. In this way, the remaining calibration error is determined by the calibration setup or, in general, by the calibration measurement accuracy. Following this procedure, the calibration setup described in this work would require $\varepsilon_{af}$ functions with five terms ($\varepsilon_{ar\_max}$ = 0.1 in Figure 7).

To study a hypothetical situation presenting lower calibration measurement errors, the fitting method is applied to the mean polarization angle error values achieved by simulation of successive calibration measurements performed with the setup of Figure 2, until getting a certain accuracy over the simulated calibration measurements. For instance, 100 measurements were simulated, achieving a final statistical measurement error of 0.1° ($1/\sqrt{100}$). In such a case, the resulting mean error values require $\varepsilon_{af}$ functions with eight or nine terms to get maximum fitting errors ($\varepsilon_{ar\_max}$) of about $10^{-2}$°. Taking this argument to an extreme (and surely not feasible) case, 10,000 calibration measurements were simulated, achieving a final accuracy of 0.01°. For such a case, $\varepsilon_{af}$ functions with 11 terms would be required to get an $\varepsilon_{ar\_max}$ of about $10^{-3}$°. Table 6 shows the $\varepsilon_{ar\_max}$ values achieved from 15 independent simulations. Ten of them considered 100 calibration measurements and the application of eight (second column) and nine (third column) $\varepsilon_{af\_i}$ terms. The last five simulations considered 10,000 calibration measurements and the application of 11 $\varepsilon_{af\_i}$ terms (fourth column).

**Table 6.** Maximum polarization angle remaining fitting errors achieved for simulated 100 (second and third column) and 10,000 (fourth column) successive measurements. Five independent simulations were performed for each number of terms (*n*).

| Simulation | $\varepsilon_{ar\_max}$ (°, *n* = 8) | $\varepsilon_{ar\_max}$ (°, *n* = 9) | $\varepsilon_{ar\_max}$ (°, *n* = 11) |
|---|---|---|---|
| 1 | $1.3 \times 10^{-2}$ | $0.65 \times 10^{-2}$ | $0.99 \times 10^{-3}$ |
| 2 | $1.6 \times 10^{-2}$ | $0.85 \times 10^{-2}$ | $1.1 \times 10^{-3}$ |
| 3 | $1.1 \times 10^{-2}$ | $0.53 \times 10^{-2}$ | $1.1 \times 10^{-3}$ |
| 4 | $0.91 \times 10^{-2}$ | $0.39 \times 10^{-2}$ | $0.93 \times 10^{-3}$ |
| 5 | $1.6 \times 10^{-2}$ | $0.73 \times 10^{-2}$ | $1.1 \times 10^{-3}$ |

As expected, the $\varepsilon_{ar\_max}$ values achieved from these simulations are very similar to those shown in Figure 7, where a calibration signal with a negligible uncertainty was considered. However, it is important to note that, in the cases of Table 4, the actual polarization angle error would be 0.1° (100 measurements) and 0.01° (10,000 measurements), which are those provided by the laboratory setup together with the realization of successive measurements, while, in Figure 7, the final error would be determined by the fitting process (a negligible calibration measurement error would be considered).

All the previous considerations can also be applied to the polarization percentage error or efficiency. Table 7 shows the evolution of the remaining errors when applying 1–5 multiplicative terms to the correction factors. It can be observed again that the error was reduced by around 50% with the application of each multiplicative term, but there were cases with lower error reduction. For instance, $O_4$ resulted in a one-term error lower than $O_1$ and $O_3$, but the final error was higher due to reductions of only 26% and 32% after applying the fourth and fifth terms, respectively. Again, how close the remaining error factors follow a sinusoidal shape can explain this behavior. In relation to the number of terms to be used in the fit, if the requirement is about 1%, only three terms are needed; however, for more demanding requirements, a higher number of terms can also be applied.

**Table 7.** Maximum polarization percentage remaining errors after application of each multiplicative term to the $fp^p_f$ functions.

| N. of Terms | $\varepsilon p^P_{max}$ (%, $O_1$) | $\varepsilon p^P_{max}$ (%, $O_2$) | $\varepsilon p^P_{max}$ (%, $O_3$) | $\varepsilon p^P_{max}$ (%, $O_4$) |
|---|---|---|---|---|
| 1 | 7.20 | 7.84 | 5.44 | 4.20 |
| 2 | 3.03 | 3.05 | 2.00 | 1.26 |
| 3 | 1.18 | 1.31 | 1.09 | 0.77 |
| 4 | 0.85 | 0.83 | 0.44 | 0.57 |
| 5 | 0.18 | 0.56 | 0.26 | 0.39 |

## 6. Conclusions

In this work, a polarization calibration method based on the use of sinusoidal fitted error functions was described and applied to a microwave polarimeter demonstrator designed to measure the CMB polarization in a frequency range from 10 to 20 GHz and based on a near-infrared (NIR) frequency up-conversion stage. The polarimeter output signals are modulated by means of an electrical phase-switching module. This modulation affects the polarization of the incoming signals, allowing their characterization. At the same time, systematic errors are also modulated, such that they can be fitted by means of sinusoidal functions. For the polarization angle calibration, the error function is composed by the sum of $n$ terms, while, for the polarization percentage, the error function is given by the product of $n$ terms, with $n$ high enough to fit the errors with the accuracy required by the particular experiment.

In an ideal case with an uncertainty in the calibration signal much lower than the remaining error values, after calibration, the polarization angle error can be reduced to the level of about 0.5° using only one or two additive terms, while reaching error levels of about 0.05° with five terms. Moreover, the polarization percentage errors are reduced to the level of about 1% using three multiplicative terms and to the level of about 0.5% with five terms. However, in a more realistic case with a calibration setup providing a given uncertainty, the applied procedure consists in fitting the errors with the required number of terms to have a remaining fitting error significantly lower than the uncertainty provided by the calibration setup. This assures the remaining calibration error to be determined by such calibration measurement uncertainty. For such a case and for a polarization angle measurement uncertainty of 1°, around five terms are required, while, for simulated uncertainties of 0.1° and 0.01°, 8–9 terms and 11 terms would be required respectively.

The addition of terms does not add significant computational effort to the systematic error correction process; however, in ultrasensitive direct imaging instruments with thousands of detectors, the proposed method should be applied to each one. In such a case, correlations between detectors should be used to reduce the computational cost and alleviate the calibration process.

The proposed method was applied to a laboratory demonstrator but can be easily applied to actual microwave polarization experiments with polarization modulation, for both ground- and space-based observatories. As the polarization angle and efficiency calibration does not require the placement of the source in the far field, this technique is suitable to be used directly in experiments such as QUIJOTE and LSPE-STRIP, as well as in others such as QUBIC, LiteBIRD, PICO, and BICEP2, assuming that the polarization errors can be characterized by their corresponding data analysis methods and provided in a format similar to that shown in this work.

**Author Contributions:** F.J.C., conceptualization of the calibration method, measurement analysis, and writing of the paper; P.V., R.B.B. and E.M.-G., setting of scientific and technical requirements, advice on microwave polarization experiments, and writing of the paper; G.P.-C., results analysis and optimization, and writing of the paper. All authors have read and agreed to the published version of the manuscript.

**Funding:** This research was funded by the Spanish Agencia Estatal de Investigación (AEI, MICIU) grant numbers ESP2017-92135-EXP, ESP2017-83921-C2-1-R, AYA2017-90675-REDC, and PID2019-110610RB-C21/AEI/10.13039/501100011033, co-funded with EU FEDER funds. It was also funded by Unidad de Excelencia María de Maeztu, grant number MDM-2017-0765.

**Informed Consent Statement:** Not applicable.

**Acknowledgments:** The authors would also like to thank D. Ortiz and the DICOM-UC group led by E. Artal for their assistance in the polarimeter demonstration and laboratory test-bench implementation.

**Conflicts of Interest:** The authors declare no conflict of interest.